\PassOptionsToPackage{unicode}{hyperref}
\PassOptionsToPackage{hyphens}{url}
\documentclass[
]{article}
\title{\texttt{gwverse}: a template for a new generic Geographically
Weighted R package}
\author{Lex Comber\textsuperscript{1}*, Chris
Brunsdon\textsuperscript{2}, Martin Callaghan\textsuperscript{1}, Paul
Harris\textsuperscript{3}, Binbin Lu\textsuperscript{4}, Nick
Malleson\textsuperscript{1}}
\date{September 2021}

\usepackage{amsmath,amssymb}
\usepackage{lmodern}
\usepackage{iftex}
\ifPDFTeX
  \usepackage[T1]{fontenc}
  \usepackage[utf8]{inputenc}
  \usepackage{textcomp} 
\else 
  \usepackage{unicode-math}
  \defaultfontfeatures{Scale=MatchLowercase}
  \defaultfontfeatures[\rmfamily]{Ligatures=TeX,Scale=1}
\fi
\IfFileExists{upquote.sty}{\usepackage{upquote}}{}
\IfFileExists{microtype.sty}{
  \usepackage[]{microtype}
  \UseMicrotypeSet[protrusion]{basicmath} 
}{}
\makeatletter
\@ifundefined{KOMAClassName}{
  \IfFileExists{parskip.sty}{%
    \usepackage{parskip}
  }{
    \setlength{\parindent}{0pt}
    \setlength{\parskip}{6pt plus 2pt minus 1pt}}
}{
  \KOMAoptions{parskip=half}}
\makeatother
\usepackage{xcolor}
\IfFileExists{xurl.sty}{\usepackage{xurl}}{} 
\IfFileExists{bookmark.sty}{\usepackage{bookmark}}{\usepackage{hyperref}}
\hypersetup{
  pdftitle={gwverse: a template for a new generic Geographically Weighted R package},
  pdfauthor={Lex Comber1*, Chris Brunsdon2, Martin Callaghan1, Paul Harris3, Binbin Lu4, Nick Malleson1},
  hidelinks,
  pdfcreator={LaTeX via pandoc}}
\usepackage[margin=1in]{geometry}
\usepackage{color}
\usepackage{fancyvrb}

\DefineVerbatimEnvironment{Highlighting}{Verbatim}{commandchars=\\\{\}}
\usepackage{framed}
\definecolor{shadecolor}{RGB}{248,248,248}
\newenvironment{Shaded}{\begin{snugshade}}{\end{snugshade}}

\newcommand{\AttributeTok}[1]{\textcolor[rgb]{0.77,0.63,0.00}{#1}}

\newcommand{\CommentTok}[1]{\textcolor[rgb]{0.56,0.35,0.01}{\textit{#1}}}

\newcommand{\ConstantTok}[1]{\textcolor[rgb]{0.00,0.00,0.00}{#1}}
\newcommand{\ControlFlowTok}[1]{\textcolor[rgb]{0.13,0.29,0.53}{\textbf{#1}}}

\newcommand{\DecValTok}[1]{\textcolor[rgb]{0.00,0.00,0.81}{#1}}
\newcommand{\DocumentationTok}[1]{\textcolor[rgb]{0.56,0.35,0.01}{\textbf{\textit{#1}}}}
\newcommand{\ErrorTok}[1]{\textcolor[rgb]{0.64,0.00,0.00}{\textbf{#1}}}

\newcommand{\FloatTok}[1]{\textcolor[rgb]{0.00,0.00,0.81}{#1}}
\newcommand{\FunctionTok}[1]{\textcolor[rgb]{0.00,0.00,0.00}{#1}}

\newcommand{\NormalTok}[1]{#1}

\newcommand{\OtherTok}[1]{\textcolor[rgb]{0.56,0.35,0.01}{#1}}

\newcommand{\SpecialCharTok}[1]{\textcolor[rgb]{0.00,0.00,0.00}{#1}}

\newcommand{\StringTok}[1]{\textcolor[rgb]{0.31,0.60,0.02}{#1}}

\usepackage{graphicx}
\makeatletter
\def\maxwidth{\ifdim\Gin@nat@width>\linewidth\linewidth\else\Gin@nat@width\fi}
\def\maxheight{\ifdim\Gin@nat@height>\textheight\textheight\else\Gin@nat@height\fi}
\makeatother
\setkeys{Gin}{width=\maxwidth,height=\maxheight,keepaspectratio}
\makeatletter
\def\fps@figure{htbp}
\makeatother
\providecommand{\tightlist}{%
  \setlength{\itemsep}{0pt}\setlength{\parskip}{0pt}}
\newlength{\cslhangindent}
\newlength{\csllabelwidth}
\newlength{\cslentryspacingunit} 
\newenvironment{CSLReferences}[2] 
 {
  \setlength{\parindent}{0pt}
  \ifodd #1
  \let\oldpar\par
  \def\par{\hangindent=\cslhangindent\oldpar}
  \fi
  \setlength{\parskip}{#2\cslentryspacingunit}
 }%
 {}
\usepackage{calc}

\usepackage{booktabs}
\usepackage{longtable}
\usepackage{array}
\usepackage{multirow}
\usepackage{wrapfig}
\usepackage{float}
\usepackage{colortbl}
\usepackage{pdflscape}
\usepackage{tabu}
\usepackage{threeparttable}
\usepackage{threeparttablex}
\usepackage[normalem]{ulem}
\usepackage{makecell}
\usepackage{xcolor}
\ifLuaTeX
  \usepackage{selnolig}  
\fi

\begin{document}
\maketitle

\textsuperscript{1} University of Leeds, UK

\textsuperscript{2} Maynooth University, Ireland

\textsuperscript{3} Rothamsted Research, UK

\textsuperscript{4} Wuhan University, China

* Email:
\href{mailto:a.comber@leeds.ac.uk}{\nolinkurl{a.comber@leeds.ac.uk}}

\hypertarget{abstract}{%
\section{Abstract}\label{abstract}}

Geographically weighted regression (GWR) is a popular approach for
investigating the spatial variation in relationships between response
and predictor variables, and critically for investigating and
understanding process spatial heterogeneity. It has been refined to
accommodate outliers, hetroskedasticity and local collinearity and
extended to LASSO and elastic net forms. The geographically weighted
(GW) framework is increasingly used to accommodate different types of
models and analyses reflecting a wider desire to explore spatial
variation in model parameters or components and to move away from
global, ``whole map'' approaches. However the growth in the use of GWR
and different GW models has only been partially supported by package
development in both R and Python, the major coding environments for
spatial analysis. The result is that refinements have been
inconsistently included (if at all) within GWR and GW functions in any
given package. As an example, the \texttt{GWmodel} R package, which
despite including the greatest number of GWR and GW related tools, has
been developed and extended on a piecemeal basis with no overarching
schema. This paper outlines the structure of a new \texttt{gwverse}
package, that will over time replace \texttt{GWmodel}, that takes
advantage of recent developments in the composition of complex,
integrated packages. It conceptualises \texttt{gwverse} as having a
modular structure, that separates core GW functionality and applications
such as GWR. It adopts a \emph{function factory} approach, in which
bespoke functions are created and returned to the user based on
user-defined parameters. Function factory approaches and the functions
they generate have the advantage of enclosing environments that are
execution environments of the function. The paper introduces and
demonstrates two modules (written as linked packages) that can be used
to undertake GWR as an initial transect through the proposed
\texttt{gwverse} schema, and to support user defined GW modules, before
discussing a number of key considerations and next steps.

\hypertarget{introduction}{%
\section{1. Introduction}\label{introduction}}

This paper describes the structure of a new over-arching R package
called \texttt{gwverse} that includes some -- but not all -- packages
for different geographically weighted tools. The aim in doing this is
two-fold. First, to re-imagine the functionality of the \texttt{GWmodel}
package (Lu et al. 2014; Gollini et al. et al. 2015) that can be used
for geographically weighted analyses of different kinds in R, including
regression. Second, and just as importantly, to include within the new
framework, structures that facilitate the development and integration of
user-defined geographically weighted tools, able to draw from the core
functionality provided by \texttt{gwverse}. The reasons for doing this
are to propose a framework that better supports users in undertaking
such analyses analyses, and critically, allows developers to easily
create and benchmark their own geographically weighted tools.

Geographically Weighted Regression (GWR, Brunsdon, Fotheringham, and
Charlton (1996)) investigates the spatial variation in relationships
between response and predictor variables. It reflects a desire to shift
away from global whole map regressions (Openshaw 1996) such as those
estimated by ordinary least squares (OLS), and including those that
account for error spatial dependence, such as regressions estimated by
restricted maximum likelihood (REML). GWR arose due to more broader
interests in investigating and understanding process spatial
heterogeneity.

\begin{figure}

{\centering \includegraphics{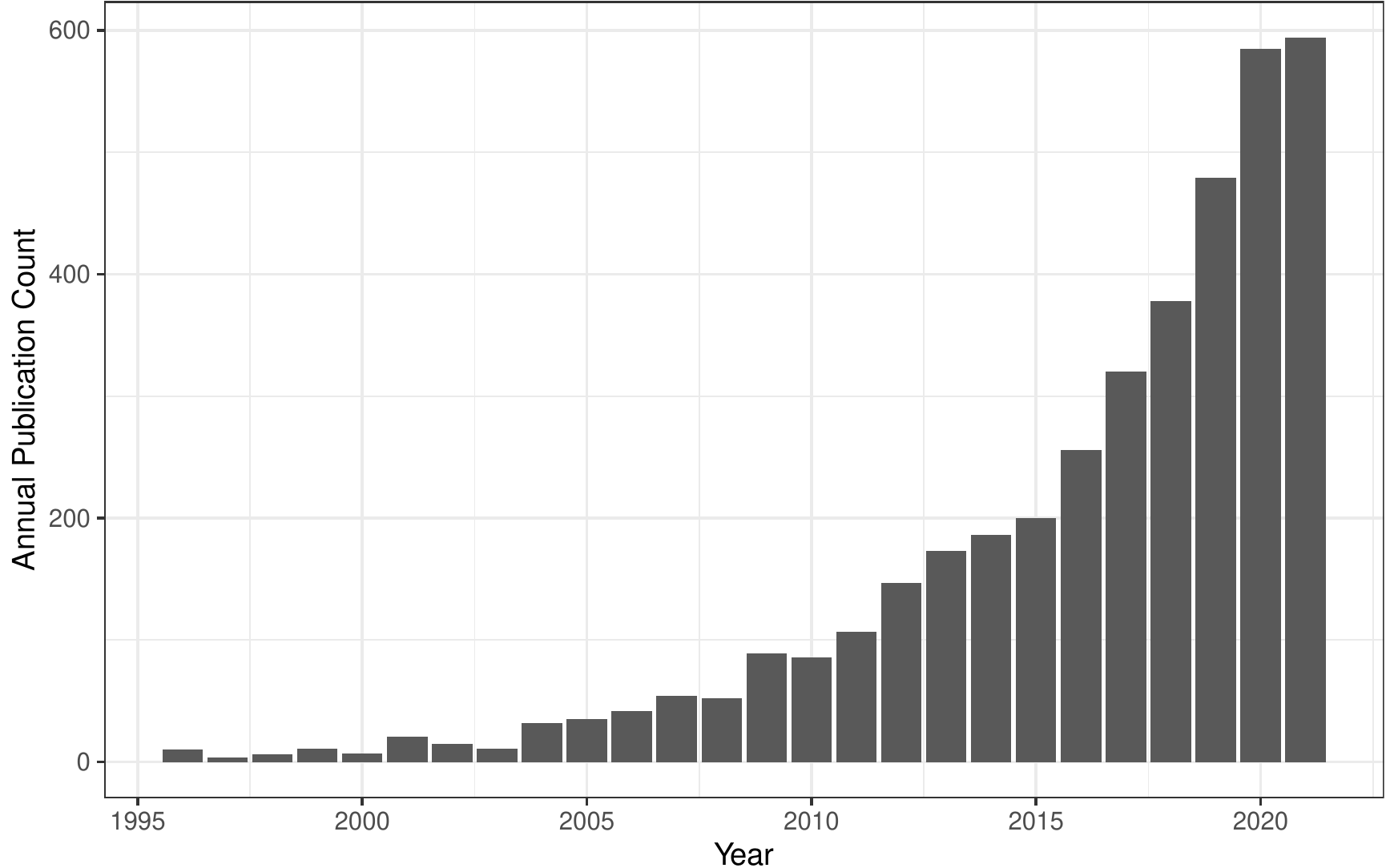} 

}

\caption{Geographically Weighted Regression publication numbers, 1996 to 2021, as listed on https://www.scopus.com.}\label{fig:fig1}
\end{figure}

GWR is increasingly being used for spatial analyses. A search of the
Scopus database (\url{https://www.scopus.com}) in September 22 2020 for
the phrases ``GWR,'' ``Geographically Weighted'' or ``Geographically
Weighted Regression'' in titles, abstracts and keywords indicated 3936
records, with sharp increases in recent years (see Figure
\ref{fig:fig1}).

This proliferation has been driven by a four main factors (Comber et al.
et al. 2020). First, is the increase in the generation, provision and
availability of spatial data (i.e.~data with some form of location
attached), and their ability to support inherently spatial analyses
(i.e.~analyses that explicitly accommodate the spatial properties of
data). Second, is a broader recognition by researchers from different
quantitative domains of the benefits of quantifying spatial patterns in
data, say through some kind of spatially informed cluster analysis or
regression technique, and in doing so handle spatial dependencies in the
data or the model parameters themselves. This has in part been driven by
the evolving adoption of the First Law of Geography as invoked by Tobler
(Tobler 1970) as a guiding principle, which in essence describes process
spatial autocorrelation (typically dependency in the data), and process
spatial heterogeneity (typically dependency in model parameters). GWR is
a method that was designed to support the latter, while will commonly
indirectly address the former (Harris 2019). Third, is the relative
simplicity and conceptual elegance of GWR as a spatial model, which has
helped to fuel its popularity. OLS regression is the basic modelling
approach (modelling 101), and the creation of local regression models
calibrated from data under a moving window, as GWR does, is conceptually
intuitive to understand. Fourth, as a result GWR has been implemented in
a number of GISs (e.g.~ESRI's ArcGIS); in R packages such as
\texttt{spgwr} (Bivand et al. 2020), \texttt{mgwrsar} (Geniaux and
Martinetti 2018), \texttt{GWLelast} (Yoneoka, Saito, and Nakaoka 2016),
\texttt{gwrr} (D. Wheeler 2013), \texttt{GWmodel} (Lu et al. 2014;
Gollini et al. et al. 2015), \texttt{McSpatial} (McMillen and McMillen
2013) and \texttt{lctools} (Kalogirou and Kalogirou 2020); in Python
packages such as \texttt{PySal} (Rey and Anselin 2010) and \texttt{mgwr}
(Oshan et al. 2019); and in standalone implementations such as
\texttt{GWR3} (Charlton, Fotheringham, and Brunsdon 2003), \texttt{GWR4}
(Nakaya et al. 2014) and \texttt{MGWR} (Z. Li et al. 2019).

GWR itself has been refined to accommodate extensions found in standard
regression, such outlier-resistant (Fotheringham, Brunsdon, and Charlton
2002;, Harris, Fotheringham, and Juggins 2010), heteroskedastic
(Fotheringham, Brunsdon, and Charlton 2002; Páez, Uchida, and Miyamoto
2002a, 2002b), ridge (D. C. Wheeler 2007; Gollini et al. et al. 2015),
LASSO (D. C. Wheeler 2009) and elastic net form (K. Li and Lam 2018;
Comber and Harris 2018). Further extensions include time in the form of
geographically and temporally weighted regression (GTWR) (Huang, Wu, and
Barry 2010; Fotheringham, Crespo, and Yao 2015), area to point
regression (Murakami and Tsutsumi 2015), multiple scales of analysis
(Yang 2014; Fotheringham, Yang, and Kang 2017), spatially variable model
specification (Comber et al. 2018) and the use of different distance
metrics (Lu et al. 2016).

A secondary tranche of developments has seen the use of the
Geographically Weighted (GW) framework as a generic structure to
accommodate different types of models and analyses. Again this reflects
a desire to explore spatial variation in model parameters or its
components and to move away from global, ``whole map'' approaches.
Examples include GW principal components analysis (PCA) (Harris,
Brunsdon, and Charlton 2011), GW descriptive statistics (Brunsdon,
Fotheringham, and Charlton 2002), GW discriminant analysis (Brunsdon,
Fotheringham, and Charlton 2007; Foley and Demšar 2013), GW
correspondence matrices and error reporting (Comber, Brunsdon, et al.
2017), GW structural equation models (Comber, Li, et al. 2017), GW
evidence combination (Comber et al. 2016), GW Variograms (Harris,
Charlton, and Fotheringham 2010), GW network design (Harris et al.
2014), GW Kriging (Harris, Charlton, and Fotheringham 2010; Harris,
Brunsdon, and Fotheringham 2011), GW visualization techniques (Dykes and
Brunsdon 2007), and more recently GW artificial neural networks (Du et
al. 2020; Hagenauer and Helbich 2021) and GW machine learning (Chen et
al. 2018; L. Li 2019; Quiñones, Goyal, and Ahmed 2021; Xu et al. 2021).
In each of these developments, the moving window or kernel is still used
to generate local data subsets as is done in GWR, that are weighted by
their distance to the kernel centre, thereby providing local inputs to
the model, analysis or evaluation being applied. These various GW models
demonstrate a generic, open, and continually evolving technical
framework that is being used to explore spatial heterogeneities from a
wide range of disciplines in the natural and social sciences.

The growth in the use of GWR and in GW models of different kinds, as
well as the refinements to GWR, has been supported to some degree by
package development in both R and Python. However, much of the
development has taken place on a piecemeal basis, extending current
functionality, without consideration of any overarching schema, nor of
more recent developments in thinking around the composition of complex,
integrated packages that incorporate a \emph{function factory} approach.
The aim of this paper is to critically examine the developments in the
package offering the greatest range of GWR and RW related functionality,
the \texttt{GWmodel} R package (Lu et al. 2014; @ Gollini et al. et al.
2015), to propose an organisational framework within which a new GWR /
GW R package will be developed, and to illustrate the first iteration of
this in a new \texttt{gwverse} R package. In so doing the paper seeks to
describe a comprehensive ecology for undertaking GWR and other GW
models, that is also able to support the generation of user-defined GW
tools.

\hypertarget{background-the-current-gwmodel-r-package}{%
\section{\texorpdfstring{2. Background: The current \texttt{GWmodel} R
package}{2. Background: The current GWmodel R package}}\label{background-the-current-gwmodel-r-package}}

\begin{figure}

{\centering \includegraphics{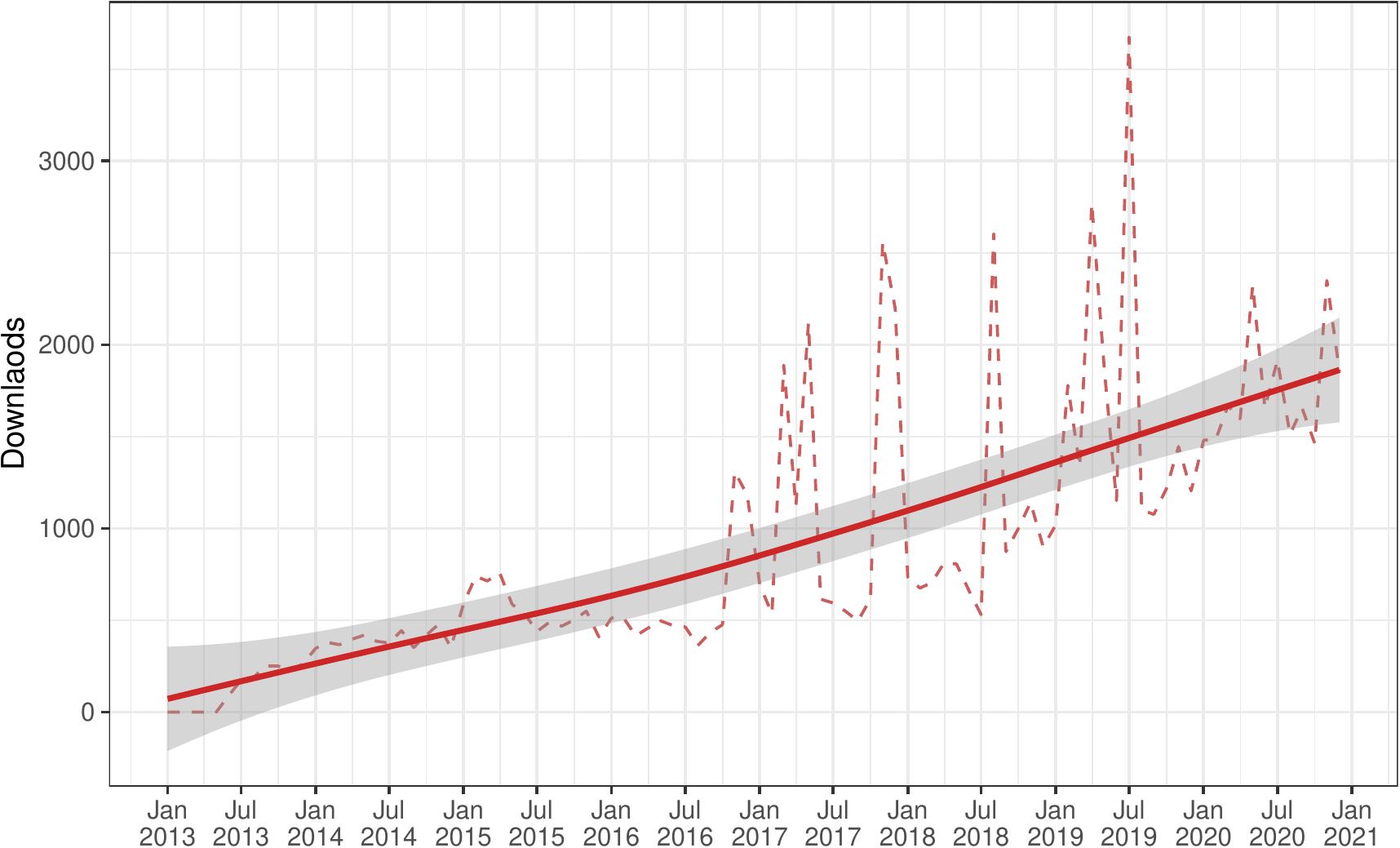} 

}

\caption{Monthly downloads of the GWmodel package from CRAN, the Comprehensive R Archive Network.}\label{fig:fig2}
\end{figure}

The \texttt{GWmodel} package provides the most comprehensive suite of
GWR and GW related tools. It contains various forms of GWR, some of
which have both basic and outlier resistant forms, some with local
statistical tests and diagnostics, a generalised linear model form, some
with options for flexible choices of distance metrics (Lu et al. 2016)
and a generalised linear model form (Fotheringham, Brunsdon, and
Charlton 2002). It also contains a number of different functions based
on the GW scheme, including tools for GW descriptive statistics, GW PCA
and GW discriminant analysis. As of September 2021, the \texttt{GWmodel}
package has been downloaded more than 176473 times since it was released
on CRAN in 2013 (as recorded on the CRAN download counts web page and
the BioConductor site). Monthly CRAN downloads are shown in Figure
\ref{fig:fig2}, indicating the increasing attraction of the package to
users from a wide range of disciplines. Additionally, the package
functionality has been constantly extended to accommodate refinements
and requests for tools from users and the package management team. For
example, modules have been incorporated over the last five years to
support GWR with large-scale data sets (Murakami et al. 2020),
multiscale GWR (Lu et al. 2017, 2018) as well as functions for GTWR and
revised algorithms for GW discriminant analysis and GW PCA.

One of the development problems, that occurs with many projects managed
by people in their spare or part-funded time, is that this growth in
package functionally mostly occurs on a piecemeal basis without the
over-arching organisation of the package being considered or revised
from its original structure. An example of this is that the
\texttt{GWmodel} manual is some 85 pages in length (Gollini et al. et
al. 2015), with varying depth of detail and two vignettes, published
with the launch of the package (Lu et al. 2014; Gollini et al. et al.
2015). The \texttt{GWmodel} help pages are similarly inconsistent. Some
functions have examples with in-depth explanations and some do not. Many
of the recent developments in package functionality do not have
vignettes, have ones that are inconsistent or have not been described in
the help pages to sufficient depth (there are also some help pages where
the example does not work). The result is that there is little to guide
the user about which functions to use and how to use them, especially
for the newer functions. This inconsistency is shown in Table
\ref{tab:tab1} which indicates the various refinements and specification
options that have been incorporated into different functions of
\texttt{GWmodel} as part of these developments, and the full description
of the package is contained in the Appendix. What is clear from Table
\ref{tab:tab1} is that whilst the complexity of the package has
increased, allowing more refined approaches to GWR for example, this
refinement has not been uniform across the main groups of GWR
functionality, or for that matter any GW model.

A further critical consideration for \texttt{GWmodel} is that currently
it only supports analyses of spatial data in \texttt{sp} format (E.
Pebesma and Bivand 2005). In the \texttt{sp} data model, spatial objects
can be thought of containing a data table of attributes and a list
structure of geometric information for the different kinds of spatial
objects (e.g.~\texttt{SpatialPointsDataFrame},
\texttt{SpatialPolygonsDataFrame} etc), where each row in the data frame
is associated with an individual component or element of the geometric
information. The \texttt{sp} class of objects is broadly analogous to
shapefile formats (lines, points, areas) and raster or grid formats.
However, \texttt{sp} is in the process of being deprecated and has been
replaced by a new class of spatial object called
\texttt{simple\ features} as implemented in the \texttt{sf} package (E.
J. Pebesma 2018). This encodes spatial data in a way that conforms to
formal standards defined in the ISO 19125-1:2004 standard. It defines a
model that allows for two-dimensional linear interpolation between
vertices and represents geometry in text (Well-known text - WKT) forms.

The \texttt{sf} structure follows a `tidy' framework (Wickham et al.
2014) and can be used with both the new native piping syntax and the
\texttt{magrittr} to undertake \texttt{dplyr} data wrangling operations.
Spatial objects in \texttt{sf} format appear as a data table but with an
extra \texttt{geometry} column that contains the WKT geometrical
information. The geometry (called an \texttt{sfc} or simple feature
column) can be used in geometric operations. Hence, due to its
complexity, over-flowing structures inconsistent application of
refinements, and the fact that it has not been revised to work with
\texttt{sf} format spatial objects, the \texttt{GWmodel} package is ripe
for an overhaul. The need for this is enhanced because of the
ever-growing popularity of GWR and the GW framework and the increasing
use of \texttt{GWmodel} to undertake these analyses. The next section
sketches out the form that this overhaul could take for a new
\texttt{gwverse} package and considers \emph{function factories} as an
approach for doing this.

\begin{table}

\caption{\label{tab:tab1}The presence of GWmodel package functionality by the main groups of related specification options (* for mean/median only), where applicable.}
\centering
\begin{tabular}[t]{l>{\raggedright\arraybackslash}p{2cm}>{\raggedright\arraybackslash}p{2cm}>{\raggedright\arraybackslash}p{2cm}>{\raggedright\arraybackslash}p{2cm}>{\raggedright\arraybackslash}p{2cm}}
\toprule{}
Option & GW Summary Statistics & GW Principal Components Analysis & GW Regression & GW Generalised Linear Models & GW Discriminant Analysis\\
\midrule{}
Flexible Distance Metric & Yes & Yes & Yes & Yes & Yes\\
Five kernel functions & Yes & Yes & Yes & Yes & Yes\\
Fixed/adaptive bandwidth & Yes & Yes & Yes & Yes & Yes\\
Bandwidth optimization & Yes* & Yes & Yes & Yes & Yes\\
Robust choice for outliers & Yes & Yes & Yes & No & No\\
Heteroskedastic errors & - & - & Yes & No & -\\
Ridge term & - & - & Yes & No & -\\
F- Tests (Leung) & - & - & Yes & - & -\\
Monte Carlo Tests & Yes & Yes & Yes & No & No\\
Bootstrap SE estimation & No & No & Yes & No & No\\
Local coefficient t-tests & - & - & Yes & Yes & No\\
Multiscale extension & - & - & Yes & No & No\\
Space-time & No & No & Yes & No & No\\
High performance & No & No & Yes & No & No\\
\bottomrule{}
\end{tabular}
\end{table}

\hypertarget{proposal-gwverse---a-template-for-a-new-gw-package}{%
\section{\texorpdfstring{3. Proposal: \texttt{gwverse} - a template for
a new GW
package}{3. Proposal: gwverse - a template for a new GW package}}\label{proposal-gwverse---a-template-for-a-new-gw-package}}

\hypertarget{the-basic-idea}{%
\subsection{3.1 The Basic Idea}\label{the-basic-idea}}

The basic idea for the \texttt{gwverse} package is to implement a
modular package structure. Such structures are seen in packages such as
\texttt{tidyverse}, which when called loads a number of
tidyverse-related packages. However, it doesn't load all
tidyverse-related packages, because this may take time, and occupy
resources. It loads more than the absolute basic \texttt{dplyr} package
(for example, it loads \texttt{ggplot}) but not \texttt{feather}. In a
similar way, we propose to have an over-arching package called
\texttt{gwverse} that loads many -- but not all -- GW-related packages.
The structure has, at its core, a package called \texttt{gw} that
provides general helper functions for a GW analysis - but essentially
provides a toolkit to be used in the construction of other GW modules.
It includes tools for building GW functions, but not the functions
themselves.

\begin{figure}

{\centering \includegraphics[width=0.9\linewidth]{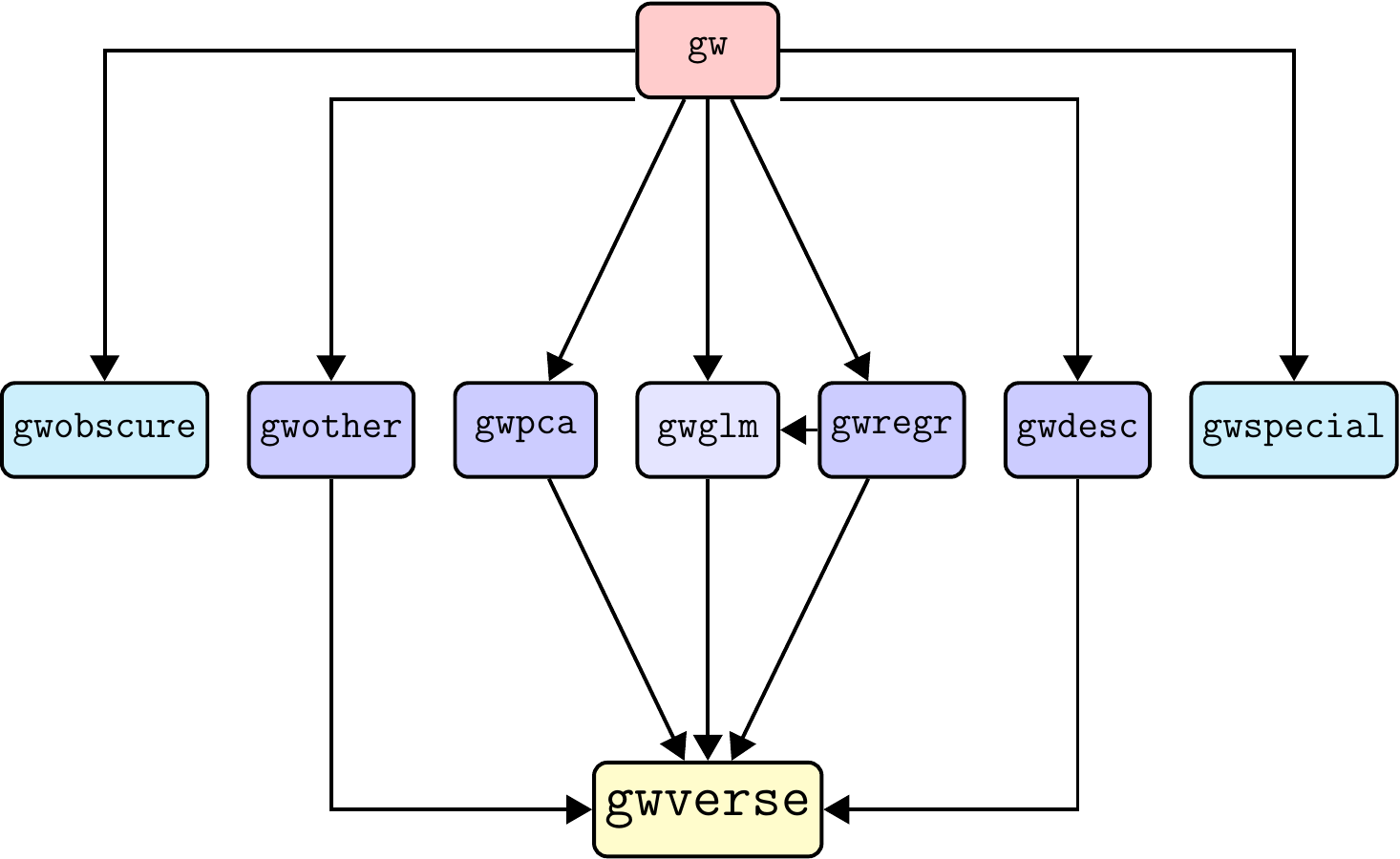} 

}

\caption{The proposed gwverse structure.}\label{fig:fig3}
\end{figure}

It is unlikely that people other than those developing GW tools will
load the \texttt{gw} package directly, rather it is implicitly loaded
(imported) when GW packages are loaded. Our proposed structure and
module dependencies are shown in Figure \ref{fig:fig3}.In this, the
boxes represent packages, and the directional arrows imply `makes use
of' or `draws functionality from.' So for example, \texttt{gwregr} (for
GWR), \texttt{gwpca} (for GW PCA) and \texttt{gwdesc} (for GW
descriptive statistics) will all use functions contained in \texttt{gw}.
The mid-layer packages are individual GW applications, and all of these
make use of \texttt{gw}. When called, the \texttt{gwverse} package loads
up several commonly used packages, although not all. The role of
\texttt{gwrglm} (for GW generalised linear models) is slightly different
as it will extensively borrow from the standard (Gaussian response) GWR
code in \texttt{gwregr}, and hence also to load and run \texttt{gwglm}
will require \texttt{gwregr} to be loaded as a dependency as well. The
packages \texttt{gwobscure} and \texttt{gwspecial} are more specialised
GW packages (as yet not identified), and so are not loaded via
\texttt{gwverse}.

In this structure, the use of \texttt{gw} as the core package is
essential to provide a consistent interface to all of the other
functions, whereas \texttt{gwverse} provides a convenient wrapper for
the modules.

\hypertarget{gw-models}{%
\subsection{3.2 GW models}\label{gw-models}}

The GW framework, for regression, or any other analysis, has at its core
a set of fundamental operations: the identification of nearby
observations to the location being considered (i.e.~observations under
the kernel) and calculation of weights for those observations based on
their distance to the location (i.e the kernel centre). The precise form
of the functions that are used to undertake these operations will depend
on user-specified choices about:

\begin{itemize}
\tightlist
\item
  kernel bandwidth type: a fixed bandwidth of a uniform size (distance),
  or an adaptive one in in which size (distance) varies but the number
  of observations is fixed or uniform;
\item
  kernel function shape: the form of the distance weighting, with
  choices of Gaussian, exponential, bisquare, tricube or boxcar and many
  more.
\end{itemize}

After the kernel bandwidth type and shape have been defined, they can be
applied to extract and weight local data in some kind of GW analysis.
For example, in GWR they are used to create a series of local
regressions (returning local coefficients and other regression related
outputs), in a GW discriminant analysis they are used to determine the
local posterior class probabilities,, in a GW PCA they are used to
determine the local components, local loadings and local scores.

The operation of a given GW model such as GWR has two stages:

\begin{enumerate}
\def\labelenumi{\arabic{enumi}.}
\tightlist
\item
  Determination of the kernel bandwidth size (whether fixed or adaptive)
  typically through some form of optimal evaluation;
\item
  Application of the optimal bandwidth in the final model.
\end{enumerate}

The nature of these stages are specific to the particular GW model and
can depend on whether or not some objective function exists (typically
whether or not the model can predict). In this respect a regression
might use a leave-one-out cross validation (CV) or an Akaike Information
Criterion (AIC) metric to evaluate different bandwidth sizes. A
discriminant analysis might use metrics commonly generated from a
correspondence analysis (matrix) such as overall classification accuracy
or the Kappa statistic.

This can be illustrated by considering the case for GWR. A GWR analysis
requires the optimal bandwidth to be determined and then used to
generate the local regressions. A linear bandwidth search would have the
following sequence, after user decisions about bandwidth type and shape:

\begin{Shaded}
\begin{Highlighting}[]
\FloatTok{1.}  \ControlFlowTok{for}\NormalTok{ each potential bandwidth}
\FloatTok{2.}  \SpecialCharTok{|}  \ControlFlowTok{for}\NormalTok{ each regression point }\SpecialCharTok{/}\NormalTok{ location}
\FloatTok{3.}  \SpecialCharTok{|}  \ErrorTok{|}\NormalTok{  identify the nearest }\StringTok{\textasciigrave{}}\AttributeTok{n}\StringTok{\textasciigrave{}}\NormalTok{ observations under the kernel}
\FloatTok{4.}  \SpecialCharTok{|}  \ErrorTok{|}\NormalTok{  calculate the observations }\FunctionTok{weights}\NormalTok{ (bisquare, Gaussian etc)}
\FloatTok{5.}  \SpecialCharTok{|}\NormalTok{  end}
\FloatTok{6.}  \SpecialCharTok{|}\NormalTok{  evlauate the fit of the local }\FunctionTok{regression}\NormalTok{ (CV or AIC) }
\FloatTok{7.}\NormalTok{  end}
\FloatTok{8.}\NormalTok{  find the best performing bandwidth via an overall CV or AIC diagnostic}
\end{Highlighting}
\end{Shaded}

Each GWR proceeds by undertaking steps 2 to 5 for the given
bandwidth\footnote{In reality a GWR evaluation by CV or AIC does not
  require the local model to be created only the observation weights at
  each location to be determined.}. For a different GW model such as GW
discriminant analysis, the outline is the broadly the same but with
different localised models in step 6 and overall evaluation in step 8.

And of course, many of these steps require a number of inputs: step 2
requires the distances between the location being considered and each
observation and as this is done for each observation suggesting the need
for a distance matrix of some kind; step 2 also requires a specific
function to identify nearby observations depending on the bandwidth
type; step 3 requires a weighting function that is specific to the
bandwidth type and shape. The point being that a number of generic
functions and data structures are required \textbf{in combination} by
any GW model, although they are used in different ways to support
different types of bandwidth evaluation and final analyses.

\hypertarget{a-function-factory-approach}{%
\subsection{3.3 A function factory
approach}\label{a-function-factory-approach}}

An alternative to the \texttt{for} loop approach above is to take a
function factory approach in combination with functionals. A
\texttt{functional} is a function that has a function as its input and
returns a vector as its output. They are commonly used alternatives to
\texttt{for} loops because they are faster\footnote{Actually they
  promote tighter coding and the avoidance of temporary data structures.}
and more flexible. A \texttt{function\ factory} is a function that
returns a function. They have the advantages of allowing values to be
precomputed within them (such as the distance matrix mentioned above),
saving computation time, of supporting a multi-level design approach
that more closely reflect the structure of the problem being addressed
(for example wrapping user defined kernel and bandwidth choices within
steps 3 and 4 above) and this way allow the complexity of the problem to
be partitioned in into more easily understood (and testable) chunks
(Wickham 2019). Examples of current R packages that take this approach
include \texttt{MCMC} (Geyer 2020)

Their key advantage is that functions generated by function factories
have an enclosing environment that is an execution environment of the
function factory. This allows, for example, the names of functions in
the enclosing environment to be associated with different function
bodies, have different values in different functions generated by
function factories (for example a CV or an AIC evolution function in the
GWR example above.) Thus the ``enclosing environment of the manufactured
function is unique and constant'' (Wickham 2019, sec. 10.2.4).

The \texttt{for} loop GWR schema presented above can be replaced with a
function factory approach, in combination with functionals, to define a
function with the \texttt{gwregr} module in Figure \ref{fig:fig3}.

First a function factory is used to create a function that returns a
function to evaluate a single bandwidth, that encloses a distance matrix
and the data needed for the analysis, along with user defined bandwidth
choices:

\begin{Shaded}
\begin{Highlighting}[]
\NormalTok{single\_bw\_gwr }\OtherTok{=} \ControlFlowTok{function}\NormalTok{(spatial\_data, adaptive, kernel\_shape, evaluation) \{}
    \DocumentationTok{\#\#\# input data related}
    \FloatTok{1.}\NormalTok{ create distance matrix from spatial\_data }
  \DocumentationTok{\#\#\# core related (i.e. functions from gw)}
    \FloatTok{2.}\NormalTok{ create the }\StringTok{\textquotesingle{}get nearby observations\textquotesingle{}} \ControlFlowTok{function}\NormalTok{ (adaptive parameter)}
    \FloatTok{3.}\NormalTok{ create the }\StringTok{\textquotesingle{}weight nearby observations\textquotesingle{}}\NormalTok{ (kernel\_shape and adaptive parameters)}
    \DocumentationTok{\#\#\# application related (i.e. function from gwregr)}
    \FloatTok{4.}\NormalTok{ define the evaluation }\ControlFlowTok{function}\NormalTok{ (evaluation parameter)}
    \DocumentationTok{\#\#\# output }
    \FloatTok{5.}\NormalTok{ define the }\ControlFlowTok{function}\NormalTok{ to be returned }
    \ControlFlowTok{function}\NormalTok{(bandwidth, spatial\_data, formula) \{}
\NormalTok{        create matrix of nearby locations }\ControlFlowTok{for}\NormalTok{ each observation }
\NormalTok{          (using the nearby }\ControlFlowTok{function}\NormalTok{, distance matrix and bandwidth);}
\NormalTok{        create matrix of weights }\ControlFlowTok{for}\NormalTok{ each observation }
\NormalTok{        (using weight }\ControlFlowTok{function}\NormalTok{, nearby locations matrix and bandwidth);}
\NormalTok{        return the results of the evaluation of the formula }
\NormalTok{          (applied to the weights matrix);}
\NormalTok{    \}}
\NormalTok{\}}
\end{Highlighting}
\end{Shaded}

This is broadly equivalent in functionality to steps 2 to 6 in the
\texttt{for} loop schema above. This function can be used to evaluate a
single bandwidth, for a given regression model equation as specified in
\texttt{formula} :

\begin{Shaded}
\begin{Highlighting}[]
\NormalTok{my\_gwr\_bandwidth\_function }\OtherTok{=} \FunctionTok{single\_bw\_gwr}\NormalTok{(}\AttributeTok{spatial\_data =}\NormalTok{ georgia, }\AttributeTok{adaptive =} \ConstantTok{TRUE}\NormalTok{, }
                                          \AttributeTok{kernel\_shape =} \StringTok{"bisquare"}\NormalTok{, }\AttributeTok{evaluation =} \StringTok{"AIC"}\NormalTok{)}
\FunctionTok{my\_gwr\_bandwidth\_function}\NormalTok{(}\AttributeTok{bw =}\DecValTok{100}\NormalTok{, georgia, formula)    }
\end{Highlighting}
\end{Shaded}

Or more commonly used to evaluate different bandwidths either using an
optimise function or through a linear search:

\begin{Shaded}
\begin{Highlighting}[]
\CommentTok{\# optimise}
\FunctionTok{optimise}\NormalTok{(my\_gwr\_bandwidth\_function, }\FunctionTok{c}\NormalTok{(}\DecValTok{10}\NormalTok{, }\FunctionTok{nrow}\NormalTok{(georgia)), }
         \AttributeTok{spatial\_data=}\NormalTok{georgia, }\AttributeTok{formula =}\NormalTok{ formula, }\AttributeTok{maximum =} \ConstantTok{FALSE}\NormalTok{)}
\CommentTok{\# linear search}
\NormalTok{bwsa }\OtherTok{=} \DecValTok{10}\SpecialCharTok{:}\FunctionTok{nrow}\NormalTok{(georgia)}
\NormalTok{res }\OtherTok{=} \FunctionTok{sapply}\NormalTok{(bwsa, }\ControlFlowTok{function}\NormalTok{(x) }\FunctionTok{my\_gwr\_bandwidth\_function}\NormalTok{(x, georgia, formula))}
\NormalTok{bwsa[}\FunctionTok{which.min}\NormalTok{(res)]}
\end{Highlighting}
\end{Shaded}

\hypertarget{gwverse-0.0.1}{%
\subsection{\texorpdfstring{3.4 \texttt{gwverse}
0.0.1}{3.4 gwverse 0.0.1}}\label{gwverse-0.0.1}}

This approach has been used to create two new packages that provide a
simple proof of concept of the new \texttt{gwverse}: the core package
\texttt{gw} and a GWR package \texttt{gwregr}. These can be installed
and used to undertake a GWR analysis in R, with fixed or adaptive
bandwidth types, with different kernel shapes, and evaluated by either
CV or corrected AIC.

The \texttt{gw} module contains three functions:

\begin{itemize}
\tightlist
\item
  \texttt{gw\_get\_nearby} which returns a function that identifies the
  observations nearby to the regression point under consideration for a
  given bandwidth. A different function is returned for adaptive and
  fixed bandwidths. The returned function takes an observation index,
  distance matrix and bandwidth as inputs and returns a vector of nearby
  observation indices.\\
\item
  \texttt{gw\_get\_weight} which returns a function for different kernel
  weights: Gaussian, bisquare, tricube, exponential and boxcar. Again, a
  different function is returned for adaptive and fixed bandwidths. The
  returned function takes a bandwidth and a vector of distances to
  nearby observations as inputs. It returns a vector of weights for
  nearby observations.
\item
  \texttt{gw\_do\_weight} which applies the selected weight function
  within an \texttt{apply} call within the function generated by the
  function factory. It takes as inputs, an index of observations, a
  bandwidth, a matrix of nearby observations for a given bandwidth, a
  distance matrix and the weight function. It returns a vector of
  weights for all observations.
\end{itemize}

The \texttt{gwregr} module contains 4 functions:

\begin{itemize}
\tightlist
\item
  \texttt{gw\_get\_lm\_eval} which returns an evaluation function to
  evaluate the GWR results for a given bandwidth. This can be specified
  as ``AIC'' or ``CV.'' The returned function takes as inputs the data
  frame of the input spatial data, a formula, the matrix of nearby
  locations and the matrix of their weights. It returns an AIC or CV
  value.
\item
  \texttt{gw\_single\_bw\_gwr} which returns a function to evaluate a
  single GWR bandwidth. It takes as input point or polygon spatial
  dataset in sf format, containing the attributes to be modelled, a
  logical value to indicate whether an adaptive or fixed bandwidth
  distance is being used, the type of distance weighting to be used, and
  evaluation method for the local model, either ``AIC'' or ``CV.'' The
  returned function generates an evaluation measure.
\item
  \texttt{gw\_do\_local\_lm} which undertakes the local weighted
  regression in an \texttt{apply} function. It takes the a vector
  weights (pertaining to an observation point), the formula and a flat
  data frame of the spatial data as inputs and returns a vector of
  coefficient estimates for the observation point being considered. This
  is not used in bandwidth selection, only in after the bandwidth has
  been specified.
\item
  \texttt{gw\_regr} which returns a function to undertake GWR once the
  optimal bandwidth has been defined. It takes as input the spatial
  dataset in sf format with the attributes to modelled, a formula, a
  logical value to indicate whether an adaptive or fixed bandwidth types
  is being used, the kernel shape and the bandwidth value. The returned
  function generates an \(n \times m\) matrix of coefficients at
  location (\(n\)) as specified in the formula (\(m\)).
\end{itemize}

These can be installed from GitHUb as follows:

\begin{Shaded}
\begin{Highlighting}[]
\FunctionTok{library}\NormalTok{(devtools)}
\FunctionTok{install\_github}\NormalTok{(}\StringTok{"gwverse/gw"}\NormalTok{)}
\FunctionTok{install\_github}\NormalTok{(}\StringTok{"gwverse/gwregr"}\NormalTok{)}
\FunctionTok{library}\NormalTok{(gwregr)}
\end{Highlighting}
\end{Shaded}

The \texttt{gwregr} package imports the \texttt{sf} package for spatial
data and comes with the well-known \texttt{georgia} dataset and this can
be loaded:

\begin{Shaded}
\begin{Highlighting}[]
\FunctionTok{data}\NormalTok{(georgia)}
\end{Highlighting}
\end{Shaded}

The first thing for GWR is bandwidth selection. A function for doing
this returned by the \texttt{gw\_single\_bw\_gwr} function and the code
below does this for an adaptive bandwidth and a bisquare kernel, applied
over the \texttt{georgia} data, using AIC as the evaluation criteria:

\begin{Shaded}
\begin{Highlighting}[]
\NormalTok{gwr\_bw\_func }\OtherTok{=} \FunctionTok{gw\_single\_bw\_gwr}\NormalTok{(georgia, }\AttributeTok{adaptive=}\ConstantTok{TRUE}\NormalTok{, }\AttributeTok{kernel=}\StringTok{"bisquare"}\NormalTok{, }\AttributeTok{eval=}\StringTok{"AIC"}\NormalTok{)}
\end{Highlighting}
\end{Shaded}

After defining a formula, the function can be run for a given bandwidth
and returns the evolution value (in this case the AIC score):

\begin{Shaded}
\begin{Highlighting}[]
\NormalTok{formula }\OtherTok{=} \FunctionTok{as.formula}\NormalTok{(MedInc }\SpecialCharTok{\textasciitilde{}}\NormalTok{ PctBach }\SpecialCharTok{+}\NormalTok{ PctEld)}
\FunctionTok{gwr\_bw\_func}\NormalTok{(}\AttributeTok{bw =}\DecValTok{100}\NormalTok{, formula)   }
\end{Highlighting}
\end{Shaded}

\begin{verbatim}
## [1] 3262.336
\end{verbatim}

Notice how no data needs to be passed to the function that is returned
as this is held in the function environment. The function environment
bindings can be explored:

\begin{Shaded}
\begin{Highlighting}[]
\FunctionTok{library}\NormalTok{(rlang)}
\CommentTok{\# environments}
\FunctionTok{env\_print}\NormalTok{(gwr\_bw\_func)}
\end{Highlighting}
\end{Shaded}

\begin{verbatim}
## <environment: 0x7ffc7c0bf1c8>
## parent: <environment: namespace:gwregr>
## bindings:
##  * eval_func: <fn>
##  * weight_func: <fn>
##  * nearby_func: <fn>
##  * df: <df[,16]>
##  * dist_mat: <dbl[,159]>
##  * adaptive: <lgl>
##  * kernel: <chr>
##  * eval: <chr>
\end{verbatim}

This indicates the objects and items that are bound to the function and
we can examine individual environment bindings such as the weighting
bi-square function:

\begin{Shaded}
\begin{Highlighting}[]
\FunctionTok{fn\_env}\NormalTok{(gwr\_bw\_func)}\SpecialCharTok{$}\NormalTok{weight\_func}
\end{Highlighting}
\end{Shaded}

\begin{verbatim}
## function (bw, dists) 
## {
##     bw = max(dists)
##     (1 - (dists/bw)^2)^2
## }
## <bytecode: 0x7ffc7bdc0898>
## <environment: 0x7ffc7bdbf698>
\end{verbatim}

or the flat data frame extracted from \texttt{georgia}:

\begin{Shaded}
\begin{Highlighting}[]
\FunctionTok{head}\NormalTok{(}\FunctionTok{fn\_env}\NormalTok{(gwr\_bw\_func)}\SpecialCharTok{$}\NormalTok{df)}
\end{Highlighting}
\end{Shaded}

\begin{verbatim}
##   Latitude  Longitud TotPop90 PctRural PctBach PctEld PctFB PctPov PctBlack
## 1 31.75339 -82.28558    15744     75.6     8.2  11.43  0.64   19.9    20.76
## 2 31.29486 -82.87474     6213    100.0     6.4  11.77  1.58   26.0    26.86
## 3 31.55678 -82.45115     9566     61.7     6.6  11.11  0.27   24.1    15.42
## 4 31.33084 -84.45401     3615    100.0     9.4  13.17  0.11   24.8    51.67
## 5 33.07193 -83.25085    39530     42.7    13.3   8.64  1.43   17.5    42.39
## 6 34.35270 -83.50054    10308    100.0     6.4  11.37  0.34   15.1     3.49
##          X       Y    ID     Name MedInc class     Inc
## 1 941396.6 3521764 13001  Appling  32152     1 low_med
## 2 895553.0 3471916 13003 Atkinson  27657     1     low
## 3 930946.4 3502787 13005    Bacon  29342     1     low
## 4 745398.6 3474765 13007    Baker  29610     4     low
## 5 849431.3 3665553 13009  Baldwin  36414     2    high
## 6 819317.3 3807616 13011    Banks  41783     3    high
\end{verbatim}

A GWR analysis can proceed with this function. The optimal bandwidth can
be determined using the \texttt{optimse} function or via a linear
search:

\begin{Shaded}
\begin{Highlighting}[]
\FunctionTok{optimise}\NormalTok{(gwr\_bw\_func, }\FunctionTok{c}\NormalTok{(}\DecValTok{10}\NormalTok{,}\FunctionTok{nrow}\NormalTok{(georgia)), }\AttributeTok{formula=}\NormalTok{formula, }\AttributeTok{maximum=}\ConstantTok{FALSE}\NormalTok{)}
\end{Highlighting}
\end{Shaded}

\begin{verbatim}
## $minimum
## [1] 48.24298
## 
## $objective
## [1] 3249.311
\end{verbatim}

A linear search of all bandwidths using a functional is slower but
confirms the search in the \texttt{optimise} function does not get lost
in local minima:

\begin{Shaded}
\begin{Highlighting}[]
\CommentTok{\# create a vector of adaptive bandwidths}
\NormalTok{bwsa }\OtherTok{=} \DecValTok{10}\SpecialCharTok{:}\FunctionTok{nrow}\NormalTok{(georgia)}
\CommentTok{\# apply the function to vector of bandwidths }
\NormalTok{res }\OtherTok{=} \FunctionTok{sapply}\NormalTok{(bwsa, }\ControlFlowTok{function}\NormalTok{(x) }\FunctionTok{gwr\_bw\_func}\NormalTok{(x, formula))}
\NormalTok{bwsa[}\FunctionTok{which.min}\NormalTok{(res)]}
\end{Highlighting}
\end{Shaded}

\begin{verbatim}
## [1] 48
\end{verbatim}

Finally a GWR analysis can be undertaken using that bandwidth:

\begin{Shaded}
\begin{Highlighting}[]
\NormalTok{bw }\OtherTok{=}\NormalTok{ bwsa[}\FunctionTok{which.min}\NormalTok{(res)]}
\NormalTok{gwr\_func }\OtherTok{=} \FunctionTok{gw\_regr}\NormalTok{(formula, georgia, }\AttributeTok{adaptive =}\NormalTok{ T, }\StringTok{"bisquare"}\NormalTok{, bw)}
\NormalTok{coef\_mat }\OtherTok{=} \FunctionTok{gwr\_func}\NormalTok{(formula)}
\FunctionTok{head}\NormalTok{(coef\_mat)}
\end{Highlighting}
\end{Shaded}

\begin{verbatim}
##          [,1]      [,2]       [,3]
## [1,] 42297.30 481.62363 -1294.0088
## [2,] 37482.48 498.04563  -931.4269
## [3,] 38650.58 567.76176 -1070.5368
## [4,] 57733.75 -94.27939 -1977.4559
## [5,] 50464.46 526.21098 -1617.5854
## [6,] 68685.45 -75.69287 -2036.6329
\end{verbatim}

And the results mapped:

\begin{Shaded}
\begin{Highlighting}[]
\FunctionTok{library}\NormalTok{(tmap)}
\FunctionTok{tm\_shape}\NormalTok{(}\FunctionTok{cbind}\NormalTok{(georgia, coef\_mat))}\SpecialCharTok{+} \FunctionTok{tm\_fill}\NormalTok{(}\StringTok{"X2"}\NormalTok{, }\AttributeTok{title =} \StringTok{"PctBach"}\NormalTok{)}
\end{Highlighting}
\end{Shaded}

\begin{figure}

{\centering \includegraphics{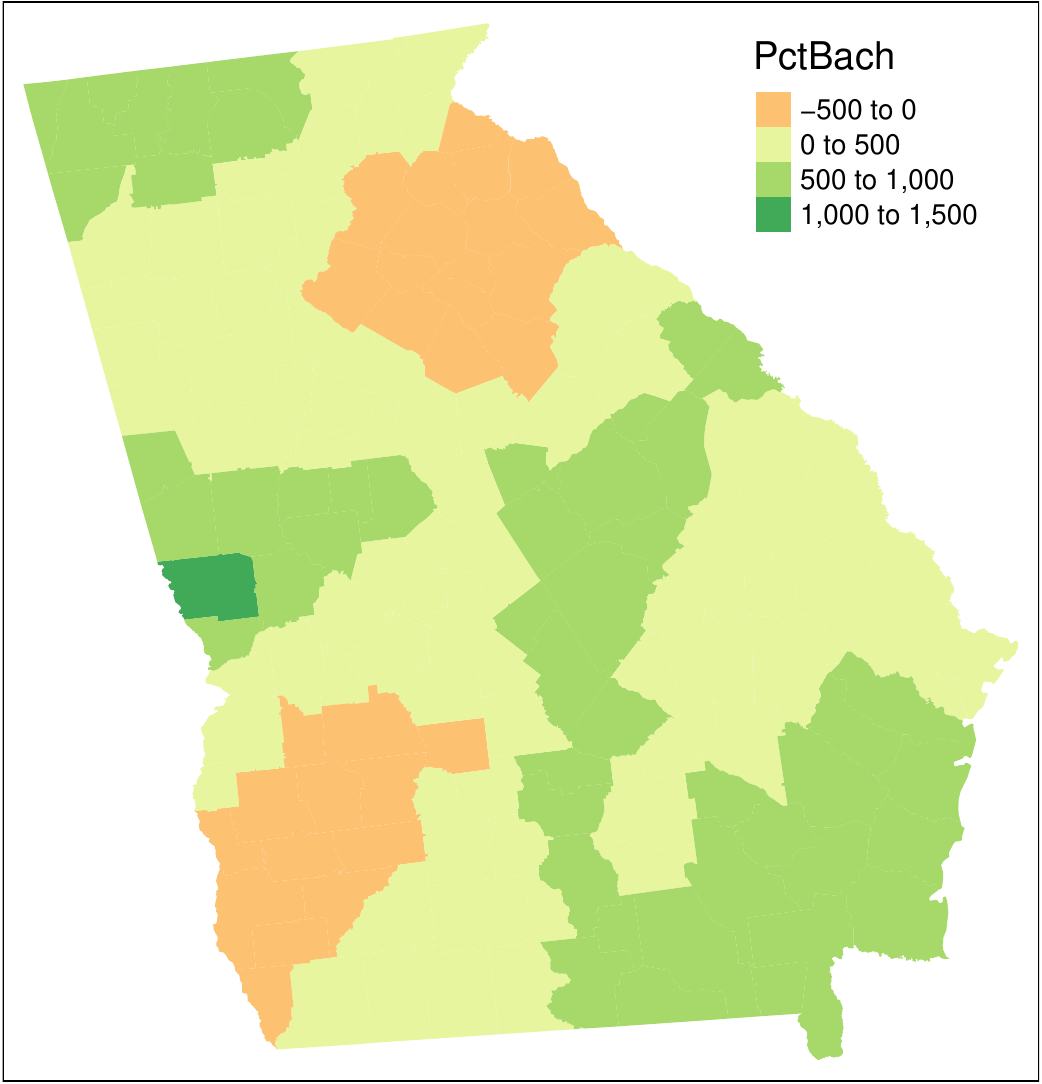} 

}

\caption{The coefficient estimates for PctBach from a GWR.}\label{fig:unnamed-chunk-20}
\end{figure}

There are a number of observations associated with this approach as
illustrated though this very simple package development:

\begin{enumerate}
\def\labelenumi{\arabic{enumi}.}
\tightlist
\item
  the \texttt{gwverse} provides a consistent framework for undertaking
  different GW analyses including GWR.
\item
  functions in the core \texttt{gw\_} module are never called directly
  by the user. Instead they are called from the modules for specific GW
  applications like GWR.
\item
  the returned functions bind what they need within their environment.
  This makes them quicker than conventional approaches despite being
  larger in working memory.
\item
  the modularisation promotes cleaner and consistent coding, allowing
  for ``all options'' of the GWR flavours in Table 1 in the future.
\item
  The idea in \texttt{gw\_} is that all of the user-end GW functions
  (e.g.~for regression, PCA, etc) are defined with this function factory
  approach - to ensure consistency of argument names (etc.) between
  functions.
\end{enumerate}

Additionally, there is a need for consistent naming conventions.
Function factories produce \textbf{anonymous functions} - i.e.~the body
of a function but unassigned to a name. The name is created when the
assignment happens. This means that although the function factory
approach guarantees consistency in interface, naming is down to
self-discipline. We suggest three naming rules: - Everything is lower
case. This is easier to use, as you don't have to remember whether a
function is called \texttt{GWModel} or \texttt{GWmodel} for example; -
Spaces in names are represented as \texttt{"\_"}; - All key functions
begin with \texttt{gw\_} - this helps on autocomplete in RStudio.

There are many potential areas of further developments such as tidy
considerations and whether functions should be pipe-able? i.e.~with the
data as the first argument, some of which are discussed below.

\hypertarget{discussion-and-conclusion}{%
\section{4. Discussion and Conclusion}\label{discussion-and-conclusion}}

Here we propose a broad framework for the development of geographically
weighted methods for spatial data analysis. Below, some implications of
this proposal will be considered. As proposers we expect to lead some of
the initial contributions -- we see development of the core \texttt{gw}
worktools, followed by further development of the package for basic GWR
(\texttt{gwregr}) as the likely first contributions, moving on to
geographically weighted generalized linear model tools, such as Poisson
and binomial regression. Other priorities include geographically
weighted descriptive statistics and PCA. We would encourage others
working in specified fields to contribute -- for example the remote
sensing community may develop GW correspondence analysis or discriminant
analysis approaches for classification, ecologists may develop
geographically weighted redundancy analysis, GW variance partitioning,
or GW canonical correspondence analysis.

The function factory approach provides a versatile framework but also
there are technical considerations. When the techniques are used,
specific `building block' functions are `bound in,' for ease of use, but
using this approach to create the techniques makes the specification of
building blocks open and explicit, as well as consistent. An alternative
may be to specify the building block functions as arguments to the main
function. However, as well as ending up with a very complex argument
list, there are issues relating to passing values to the building block
functions, as well as handling the R dot-dot-dot (\texttt{...})
parameter syntax and partial parameter name matches (Geyer 2020). There
are, however some potential problems which must be considered. For
example, when binding very large environments to a function (such as a
large spatial database), this involves making a copy of that database --
which could lead to storage or memory issues. There is a need for a set
of guidelines on good practice for the use of function factories.

There are issues in code development to be addressed. For example in the
\texttt{GWmodel} package, extensive use of linking R to C++ was made, to
enable code to run faster. This would also be useful in this proposal --
but a consistent approach to incorporating compiled C++ routines would
be needed, and in addition to a GW core package (\texttt{gw}) providing
R tools, a similar set of core procedures in C++ may be needed. In
addition, many users and developers of geographically weighted methods
use Python rather than R. Some consideration of interoperability could
allow some degree of collaboration -- working together to some extent
could become possible via the use of the \texttt{reticulate} library,
which allows Python code to run within R. A longer term goal may be to
provide a suite of Python modules mirroring the \texttt{gwverse}
approach, encouraging the same development framework to run in parallel
for the R and Python user communities.

Also, interoperability between the \texttt{gwverse} approach and other R
packages and package families can be considered. For example, many users
are now trained to work primarily with pipeline operators in R (either
the \texttt{\%\textgreater{}\%} operator from \texttt{magrittr}, or the
native \texttt{\textbar{}\textgreater{}} operator) and designing gwverse
functions to combine simply and intuitively with tools from other
packages is an important consideration. This involves thought about
which argument in \texttt{gwverse} functions should be first, and what
form the returned value of the functions take. Ideally, one would wish
\texttt{gwverse} functions to combine easily with functions from
\texttt{sf} and data manipulation tools from \texttt{tidyverse}.

Some organizational issues also require consideration. In particular we
are proposing a family of R packages, maintained on GitHub, with
periodic updates to CRAN. Although the approach here involves -- and
indeed encourages -- collaboration, the project will require curation
(much as CRAN does, but on a much smaller scale), and some structure for
this needs to be agreed. This needs to address standards for package
\texttt{gwverse} contribution -- for example, checking that the package
properly meets the function naming requirements set out above, and
assessing whether any attached vignettes are well written and with
sufficient content, and checking whether CRAN's requirements are met
when versions are submitted there. There may also be issues of managing
contributions -- for example if two contributors simultaneously propose
packages for the same (or at least overlapping) GW techniques. Another
related issue may be to create guidelines for developers, possibly
combined with a `how-to' manual outlining the use of GitHub, and the
agreed curation framework. In this way users who have a question with no
current means of investigation could be encouraged to become developers.

In conclusion, in this paper we have presented a framework for
developing a consistent and interoperable family of R packages for
geographically weighted analytical methods. We advocate the use of
functionals and function factories as key principles in this framework.
This offers a number of advantages: it facilitates the creation of
packages having a consistent interface -- so that for example an
argument to a function specifying bandwidth always has the same name and
the data supplied always has the same format. In addition, the use of a
GW core package ensures that procedures appearing in several
geographically weighted methods do not have to be re-created repeatedly
(and possibly inconsistently) in several packages. With current trends
suggesting an increasing interest in a number of new approaches, such as
space-time weighting and multiscale models, perhaps now is an
advantageous time to provide a consistent set of tools for software
development. Potentially this brings together disparate GW working
groups (of both users and developers) together under the same framework
with mutual benefit to all, allowing rapid development of new GW models
with a more streamlined community review process.

\newpage

\hypertarget{appendix-the-current-structure-of-the-gwmodel-package}{%
\section{\texorpdfstring{Appendix: the current structure of the
\texttt{GWmodel}
package}{Appendix: the current structure of the GWmodel package}}\label{appendix-the-current-structure-of-the-gwmodel-package}}

\hypertarget{overview}{%
\subsection{Overview}\label{overview}}

\texttt{GWmodel} includes functions to calibrate and estimate a wide
range of techniques based on geographical weighting. These include:
summary statistics, principal components analysis, discriminant analysis
and various forms of regression; some of which are provided in basic and
outlier resistant forms.

However the manual is, at the time of writing is long, some 85 pages in
length. It is also organised alphabetically, and while the write-ups
conform to the CRAN guidelines, they can be hard to follow. For some of
the more complex techniques there is little to guide the user as to
which functions to use. This appendix provides a structured overview of
the \texttt{GWmodel} library. It has been divided into sections, each
section containing a group of related functions. Each section is headed
with a summary table, giving the name of each function and a one-line
description of its action. Below each table a bulleted list containing
slightly more extended, but brief, descriptions of the function.

\begin{figure}

{\centering \includegraphics[width=0.9\linewidth]{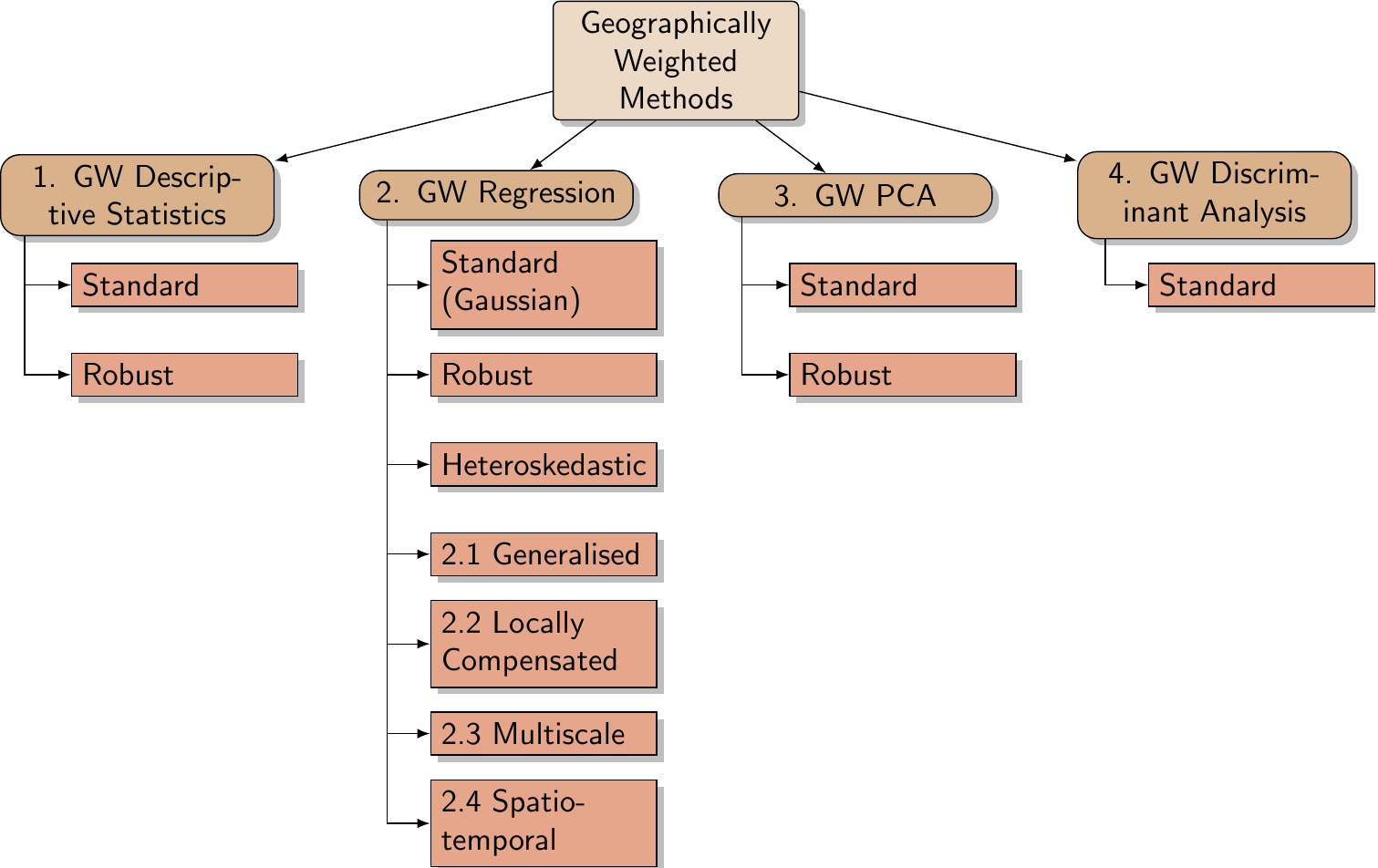} 

}

\caption{An  Ontology of GW Approaches Currently in GWmodel.}\label{fig:unnamed-chunk-21}
\end{figure}

\hypertarget{datasets}{%
\subsection{Datasets}\label{datasets}}

There are several built-in datasets. Most of these are used in the
example code which is provided at the end of each function description.
However, you can also use them to practice on, or as test data in their
own right. With the exception of \texttt{Georgia} these are all in
Spatial Polygon or Spatial Point Data Frame \texttt{sp} formats.

\begin{itemize}
\tightlist
\item
  \texttt{Dubvoter}: Voter turnout and social characters data in Greater
  Dublin for the 2002 General election and the 2002 census. Note that
  this data set was originally thought to relate to 2004, so for
  continuity we have retained the associated variable names.
\item
  \texttt{EWHP}: A house price data set for England and Wales from 2001
  with 9 hedonic (explanatory) variables.
\item
  \texttt{EWOutline}: Outline (SpatialPolygonsDataFrame) of the England
  and Wales house price data EWHP
\item
  \texttt{Georgia}: Census data from the county of Georgia, USA
\item
  \texttt{GeorgiaCounties}: The Georgia census data with boundaries for
  mapping
\item
  \texttt{LondonBorough}: Outline (SpatialPolygonsDataFrame) of London
  boroughs for the \texttt{LondonHP} data.
\item
  \texttt{LondonHP}: A house price data set with 18 hedonic variables
  for London in 2001.
\item
  \texttt{USelect}: Results of the 2004 US presidential election at the
  county level, together with five socio-economic (census) variables.
  This data can be used with GW Discriminant Analysis.
\end{itemize}

\hypertarget{service-functions}{%
\subsection{Service functions}\label{service-functions}}

\begin{itemize}
\tightlist
\item
  \texttt{gw.dist}: Calculate a distance vector(matrix) between any GW
  model calibration point(s) and the data points.
\item
  \texttt{gw.weight}: Calculate a weight vector(matrix) from a distance
  vector(matrix).
\item
  \texttt{gwr.write}: This function writes the calibration result of
  function \texttt{gwr.basic} to a text file
\item
  \texttt{gwr.write.shp}: This function writes the calibration result of
  function \texttt{gwr.basic} to a shapefile
\end{itemize}

\hypertarget{gw-descriptive-statistics}{%
\subsection{1. GW Descriptive
statistics}\label{gw-descriptive-statistics}}

\begin{itemize}
\tightlist
\item
  \texttt{bw.gwss.average}: A function for automatic bandwidth
  selections to calculate GW summary averages, including means and
  medians, via a cross-validation approach.
\item
  \texttt{gwss}: This function calculates basic and robust GWSS. This
  includes geographically weighted means, standard deviations and skew.
  Robust alternatives include geographically weighted medians,
  interquartile ranges and quantile imbalances. This function also
  calculates basic geographically weighted covariances together with
  basic and robust geographically weighted correlations.
\item
  \texttt{gwss.montecarlo}: This function implements Monte Carlo
  (randomisation) tests for the GW summary statistics found in
  \texttt{gwss}.
\item
  \texttt{gw.pcplot}: This function provides a geographically weighted
  parallel coordinate plot for locally investigating a multivariate data
  set. It has an option that weights the lines of the plot with
  increasing levels of transparency, according to their observation's
  distance from a specified focal/observation point.
\item
  \texttt{gwpca.glyph.plot}: This function provides a multivariate glyph
  plot of GWPCA loadings at each output location.
\end{itemize}

\hypertarget{gw-gaussian-regression}{%
\subsection{2. GW (gaussian) Regression}\label{gw-gaussian-regression}}

\begin{itemize}
\tightlist
\item
  \texttt{bw.gwr}: A function for automatic bandwidth selection to
  calibrate a basic GWR model.
\item
  \texttt{gwr.basic}: This function implements basic GWR.
\item
  \texttt{gwr.robust}: This function implements two robust GWR models.
\item
  \texttt{gwr.hetero}: This function implements a heteroskedastic GWR
  model
\item
  \texttt{gwr.bootstrap}: This function implements bootstrap methods to
  test for coefficients.
\item
  \texttt{gwr.montecarlo}: This function implements a Monte Carlo
  (randomisation) test to test for significant (spatial) variability of
  a GWR model's parameters or coefficients.
\item
  \texttt{gwr.t.adjust}: Given a set of p-values from the pseudo t-tests
  of basic GWR outputs, this function returns adjusted p-values using:
  (a) Bonferroni, (b) Benjamini-Hochberg, (c) Benjamini-Yekutieli and
  (d) Fotheringham-Byrne procedures.
\item
  \texttt{gwr.model.selection}: This function selects one GWR model from
  many alternatives based on the AICc values.
\item
  \texttt{gwr.model.sort}: Sort the results from the GWR model selection
  function \texttt{gwr.model.selection}.
\item
  \texttt{gwr.model.view}: This function visualises the GWR models from
  \texttt{gwr.model.selection}.
\item
  \texttt{gwr.mink.approach}: This function implements the Minkowski
  approach to select an optimum distance metric for calibrating a GWR
  model.
\item
  \texttt{gwr.mink.matrixview}: This function visualises the AICc/CV
  results from the \texttt{gwr.mink} approach.
\item
  \texttt{gwr.mink.pval}: These functions implement heuristics to select
  the values of p from two intervals: \((0,2]\) in a backward direction
  and \((2, \infty)\) in a forward direction.
\end{itemize}

\hypertarget{generalsed-gwr}{%
\subsubsection{2.1 Generalsed GWR}\label{generalsed-gwr}}

\begin{itemize}
\tightlist
\item
  \texttt{bw.ggwr}: A function for automatic bandwidth selection to
  calibrate a generalised GWR model.
\item
  \texttt{ggwr.basic}: This function implements generalised GWR.
\item
  \texttt{ggwr.cv}: This function finds the cross-validation score for a
  specified bandwidth for generalised GWR. It can be used to construct
  the bandwidth function across all possible bandwidths and compared to
  that found automatically.
\item
  \texttt{ggwr.cv.contrib}: This function finds the individual
  cross-validation score at each observation location, for a generalised
  GWR model, for a specified bandwidth. These data can be mapped to
  detect unusually high or low cross-validations scores.
\end{itemize}

\hypertarget{locally-compensated-regression}{%
\subsubsection{2.2 Locally compensated
regression}\label{locally-compensated-regression}}

\begin{itemize}
\tightlist
\item
  \texttt{bw.gwr.lcr}: A function for automatic bandwidth selection for
  \texttt{gwr.lcr} via a cross-validation approach only.
\item
  \texttt{gwr.lcr}: To address possible local collinearity problems in
  basic GWR, GWR-LCR finds local ridge parameters at affected locations
  (set by a user-specified threshold for the design matrix condition
  number).
\item
  \texttt{gwr.lcr.cv}: This function finds the cross-validation score
  for a specified bandwidth for GWR-LCR. It can be used to construct the
  bandwidth function across all possible bandwidths and compared to that
  found automatically.
\item
  \texttt{gwr.lcr.cv.contrib}: This function finds the individual
  cross-validation score at each observation location, for a GWRLCR
  model, for a specified bandwidth. These data can be mapped to detect
  unusually high or low cross-validations scores.
\item
  \texttt{gwr.collin.diagno}: This function provides a series of local
  collinearity diagnostics for the independent variables of a basic GWR
  model.
\end{itemize}

\hypertarget{multiscale-gwr}{%
\subsubsection{2.3 Multiscale GWR}\label{multiscale-gwr}}

\begin{itemize}
\tightlist
\item
  \texttt{gwr.mixed}: This function implements mixed (semi-parametric)
  GWR.
\item
  \texttt{gwr.multiscale}: This function implements multiscale GWR to
  detect variations in regression relationships across different spatial
  scales. This function can not only find a different bandwidth for each
  relationship but also (and simultaneously) find a different distance
  metric for each relationship (if required to do so).
\end{itemize}

\hypertarget{geographically-and-temporally-weighted-regression}{%
\subsubsection{2.4 Geographically and temporally weighted
regression}\label{geographically-and-temporally-weighted-regression}}

\begin{itemize}
\tightlist
\item
  \texttt{bw.gtwr}: A function for automatic bandwidth selection to
  calibrate a Geographically and Temporally Weighted Regression (GTWR)
  model.
\item
  \texttt{gtwr}: A function for calibrating a GTWR model.
\end{itemize}

\hypertarget{geographically-weighted-principal-components-analysis}{%
\subsection{3. Geographically weighted principal components
analysis}\label{geographically-weighted-principal-components-analysis}}

\begin{itemize}
\tightlist
\item
  \texttt{bw.gwpca}: A function for automatic bandwidth selection to
  calibrate a basic or robust GWPCA via a cross-validation approach only
\item
  \texttt{gwpca}: This function implements basic or robust GWPCA.
\item
  \texttt{gwpca.check.components}: The function interacts with the
  multivariate glyph plot of GWPCA loadings.
\item
  \texttt{gwpca.cv}: This function finds the cross-validation score for
  a specified bandwidth for basic or robust GWPCA. It can be used to
  construct the bandwidth function across all possible bandwidths and
  compared to that found automatically.
\item
  \texttt{gwpca.cv.contrib}: This function finds the individual
  cross-validation score at each observation location, for a GWPCA
  model, for a specified bandwidth. These data can be mapped to detect
  unusually high or low cross-validation scores
\item
  \texttt{gwpca.montecarlo.1}: This function implements a Monte Carlo
  (randomisation) test for a basic or robust GW PCA with the bandwidth
  pre-specified and constant. The test evaluates whether the GW
  eigenvalues vary significantly across space for the first component
  only.
\item
  \texttt{gwpca.montecarlo.2}: This function implements a Monte Carlo
  (randomisation) test for a basic or robust GW PCA with the bandwidth
  automatically re-selected via the cross-validation approach. The test
  evaluates whether the GW eigenvalues vary significantly across space
  for the first component only.
\end{itemize}

\hypertarget{geographically-weighted-discriminant-analysis}{%
\subsection{4. Geographically weighted discriminant
analysis}\label{geographically-weighted-discriminant-analysis}}

\begin{itemize}
\tightlist
\item
  \texttt{bw.gwda}: A function for automatic bandwidth selection for GW
  Discriminant Analysis using a cross-validation approach only
\item
  \texttt{gwda}: A function to implement GW discriminant analysis.
\end{itemize}

\hypertarget{references}{%
\section*{References}\label{references}}
\addcontentsline{toc}{section}{References}

\hypertarget{refs}{}
\begin{CSLReferences}{1}{0}
\leavevmode\vadjust pre{\hypertarget{ref-bivand2020package}{}}%
Bivand, Roger, Danlin Yu, Tomoki Nakaya, Miquel-Angel Garcia-Lopez, and
Maintainer Roger Bivand. 2020. {``Package {`Spgwr'}.''} \emph{R Software
Package}.

\leavevmode\vadjust pre{\hypertarget{ref-brunsdon1996geographically}{}}%
Brunsdon, Chris, A Stewart Fotheringham, and Martin E Charlton. 1996.
{``Geographically Weighted Regression: A Method for Exploring Spatial
Nonstationarity.''} \emph{Geographical Analysis} 28 (4): 281--98.

\leavevmode\vadjust pre{\hypertarget{ref-brunsdon2002geographically}{}}%
Brunsdon, Chris, AS Fotheringham, and Martin Charlton. 2002.
{``Geographically Weighted Summary Statistics---a Framework for
Localised Exploratory Data Analysis.''} \emph{Computers, Environment and
Urban Systems} 26 (6): 501--24.

\leavevmode\vadjust pre{\hypertarget{ref-brunsdon2007geographically}{}}%
Brunsdon, Chris, Stewart Fotheringham, and Martin Charlton. 2007.
{``Geographically Weighted Discriminant Analysis.''} \emph{Geographical
Analysis} 39 (4): 376--96.

\leavevmode\vadjust pre{\hypertarget{ref-charlton2003}{}}%
Charlton, Martin, AS Fotheringham, and Chris Brunsdon. 2003. {``GWR 3:
Software for Geographically Weighted Regression.''} Nattional Centre for
GeoComputation, National University of Ireland Maynooth.

\leavevmode\vadjust pre{\hypertarget{ref-chen2018estimation}{}}%
Chen, Lin, Chunying Ren, Bai Zhang, Zongming Wang, and Yanbiao Xi. 2018.
{``Estimation of Forest Above-Ground Biomass by Geographically Weighted
Regression and Machine Learning with Sentinel Imagery.''} \emph{Forests}
9 (10): 582.

\leavevmode\vadjust pre{\hypertarget{ref-comber2020gwr}{}}%
Comber, Alexis, Chris Brunsdon, Martin Charlton, Guanpeng Dong, Rich
Harris, Binbin Lu, Yihe Lü, et al., et al. 2020. {``The GWR Route Map: A
Guide to the Informed Application of Geographically Weighted
Regression.''} \emph{arXiv Preprint arXiv:2004.06070}.

\leavevmode\vadjust pre{\hypertarget{ref-comber2017geographically}{}}%
Comber, Alexis, Chris Brunsdon, Martin Charlton, and Paul Harris. 2017.
{``Geographically Weighted Correspondence Matrices for Local Error
Reporting and Change Analyses: Mapping the Spatial Distribution of
Errors and Change.''} \emph{Remote Sensing Letters} 8 (3): 234--43.

\leavevmode\vadjust pre{\hypertarget{ref-comber2016geographically}{}}%
Comber, Alexis, Cidália Fonte, Giles Foody, Steffen Fritz, Paul Harris,
Ana-Maria Olteanu-Raimond, and Linda See. 2016. {``Geographically
Weighted Evidence Combination Approaches for Combining Discordant and
Inconsistent Volunteered Geographical Information.''}
\emph{GeoInformatica} 20 (3): 503--27.

\leavevmode\vadjust pre{\hypertarget{ref-comber2018geographically}{}}%
Comber, Alexis, and Paul Harris. 2018. {``Geographically Weighted
Elastic Net Logistic Regression.''} \emph{Journal of Geographical
Systems} 20 (4): 317--41.

\leavevmode\vadjust pre{\hypertarget{ref-comber2017gwsem}{}}%
Comber, Alexis, Ting Li, Yihe Lü, Bojie Fu, and Paul Harris. 2017.
{``Geographically Weighted Structural Equation Models: Spatial Variation
in the Drivers of Environmental Restoration Effectiveness.''} In
\emph{Societal Geo-Innovation. 20th AGILE Conference Proceedings}.

\leavevmode\vadjust pre{\hypertarget{ref-comber2018hyper}{}}%
Comber, Alexis, Yunqiang Wang, Yihe Lü, Xingchang Zhang, and Paul
Harris. 2018. {``Hyper-Local Geographically Weighted Regression:
Extending GWR Through Local Model Selection and Local Bandwidth
Optimization.''} \emph{Journal of Spatial Information Science}, no. 17:
63--84.

\leavevmode\vadjust pre{\hypertarget{ref-du2020geographically}{}}%
Du, Zhenhong, Zhongyi Wang, Sensen Wu, Feng Zhang, and Renyi Liu. 2020.
{``Geographically Neural Network Weighted Regression for the Accurate
Estimation of Spatial Non-Stationarity.''} \emph{International Journal
of Geographical Information Science} 34 (7): 1353--77.

\leavevmode\vadjust pre{\hypertarget{ref-dykes2007geographically}{}}%
Dykes, Jason, and Chris Brunsdon. 2007. {``Geographically Weighted
Visualization: Interactive Graphics for Scale-Varying Exploratory
Analysis.''} \emph{IEEE Transactions on Visualization and Computer
Graphics} 13 (6): 1161--68.

\leavevmode\vadjust pre{\hypertarget{ref-foley2013using}{}}%
Foley, Peter, and Urška Demšar. 2013. {``Using Geovisual Analytics to
Compare the Performance of Geographically Weighted Discriminant Analysis
Versus Its Global Counterpart, Linear Discriminant Analysis.''}
\emph{International Journal of Geographical Information Science} 27 (4):
633--61.

\leavevmode\vadjust pre{\hypertarget{ref-fotheringham2003geographically}{}}%
Fotheringham, A Stewart, Chris Brunsdon, and Martin Charlton. 2002.
\emph{Geographically Weighted Regression: The Analysis of Spatially
Varying Relationships}. John Wiley \& Sons.

\leavevmode\vadjust pre{\hypertarget{ref-fotheringham2015geographical}{}}%
Fotheringham, A Stewart, Ricardo Crespo, and Jing Yao. 2015.
{``Geographical and Temporal Weighted Regression (GTWR).''}
\emph{Geographical Analysis} 47 (4): 431--52.

\leavevmode\vadjust pre{\hypertarget{ref-fotheringham2017multiscale}{}}%
Fotheringham, A Stewart, Wenbai Yang, and Wei Kang. 2017. {``Multiscale
Geographically Weighted Regression (MGWR).''} \emph{Annals of the
American Association of Geographers} 107 (6): 1247--65.

\leavevmode\vadjust pre{\hypertarget{ref-geniaux2018new}{}}%
Geniaux, Ghislain, and Davide Martinetti. 2018. {``A New Method for
Dealing Simultaneously with Spatial Autocorrelation and Spatial
Heterogeneity in Regression Models.''} \emph{Regional Science and Urban
Economics} 72: 74--85.

\leavevmode\vadjust pre{\hypertarget{ref-geyer2020mcmc}{}}%
Geyer, Charles J. 2020. {``MCMC Package Example (Version 0.9.7).''}
\url{https://cran.rediris.es/web/packages/mcmc/vignettes/demo.pdf}.

\leavevmode\vadjust pre{\hypertarget{ref-gollini2015gwmodel}{}}%
Gollini, Isabella, Binbin Lu, Martin Charlton, Christopher Brunsdon,
Paul Harris et al. 2015. {``GWmodel: An r Package for Exploring Spatial
Heterogeneity Using Geographically Weighted Models.''} \emph{Journal of
Statistical Software} 63 (i17).

\leavevmode\vadjust pre{\hypertarget{ref-hagenauer2021geographically}{}}%
Hagenauer, Julian, and Marco Helbich. 2021. {``A Geographically Weighted
Artificial Neural Network.''} \emph{International Journal of
Geographical Information Science}, 1--21.

\leavevmode\vadjust pre{\hypertarget{ref-harris2019simulation}{}}%
Harris, Paul. 2019. {``A Simulation Study on Specifying a Regression
Model for Spatial Data: Choosing Between Autocorrelation and
Heterogeneity Effects.''} \emph{Geographical Analysis} 51 (2): 151--81.
https://doi.org/\url{https://doi.org/10.1111/gean.12163}.

\leavevmode\vadjust pre{\hypertarget{ref-harris2011geographically}{}}%
Harris, Paul, Chris Brunsdon, and Martin Charlton. 2011.
{``Geographically Weighted Principal Components Analysis.''}
\emph{International Journal of Geographical Information Science} 25
(10): 1717--36.

\leavevmode\vadjust pre{\hypertarget{ref-harris2011links}{}}%
Harris, Paul, Chris Brunsdon, and A Stewart Fotheringham. 2011.
{``Links, Comparisons and Extensions of the Geographically Weighted
Regression Model When Used as a Spatial Predictor.''} \emph{Stochastic
Environmental Research and Risk Assessment} 25 (2): 123--38.

\leavevmode\vadjust pre{\hypertarget{ref-harris2010moving}{}}%
Harris, Paul, Martin Charlton, and A Stewart Fotheringham. 2010.
{``Moving Window Kriging with Geographically Weighted Variograms.''}
\emph{Stochastic Environmental Research and Risk Assessment} 24 (8):
1193--1209.

\leavevmode\vadjust pre{\hypertarget{ref-harris2014geographically}{}}%
Harris, Paul, Annemarie Clarke, Steve Juggins, Chris Brunsdon, and
Martin Charlton. 2014. {``Geographically Weighted Methods and Their Use
in Network Re-Designs for Environmental Monitoring.''} \emph{Stochastic
Environmental Research and Risk Assessment} 28 (7): 1869--87.

\leavevmode\vadjust pre{\hypertarget{ref-harris2010robust}{}}%
Harris, Paul, A Stewart Fotheringham, and Steve Juggins. 2010. {``Robust
Geographically Weighted Regression: A Technique for Quantifying Spatial
Relationships Between Freshwater Acidification Critical Loads and
Catchment Attributes.''} \emph{Annals of the Association of American
Geographers} 100 (2): 286--306.

\leavevmode\vadjust pre{\hypertarget{ref-huang2010geographically}{}}%
Huang, Bo, Bo Wu, and Michael Barry. 2010. {``Geographically and
Temporally Weighted Regression for Modeling Spatio-Temporal Variation in
House Prices.''} \emph{International Journal of Geographical Information
Science} 24 (3): 383--401.

\leavevmode\vadjust pre{\hypertarget{ref-kalogirou2020package}{}}%
Kalogirou, Stamatis, and Maintainer Stamatis Kalogirou. 2020. {``Package
{`Lctools'}.''} \emph{Local Correlation, Spatial Inequalities and Other
Tools. Available Online: Https://CRAN. R-Project. Org/Package= Lctools
(Accessed on 15 September 2020)}.

\leavevmode\vadjust pre{\hypertarget{ref-li2018geographically}{}}%
Li, Kenan, and Nina SN Lam. 2018. {``Geographically Weighted Elastic
Net: A Variable-Selection and Modeling Method Under the Spatially
Nonstationary Condition.''} \emph{Annals of the American Association of
Geographers} 108 (6): 1582--1600.

\leavevmode\vadjust pre{\hypertarget{ref-li2019geographically}{}}%
Li, Lianfa. 2019. {``Geographically Weighted Machine Learning and
Downscaling for High-Resolution Spatiotemporal Estimations of Wind
Speed.''} \emph{Remote Sensing} 11 (11): 1378.

\leavevmode\vadjust pre{\hypertarget{ref-limgwr}{}}%
Li, Ziqi, Taylor Oshan, Stewart Fotheringham, Wei Kang, Levi Wolf,
Hanchen Yu, and Wei Luo. 2019. {``MGWR.''} Spatial Analysis Research
Center, Arizona State University.

\leavevmode\vadjust pre{\hypertarget{ref-lu2017geographically}{}}%
Lu, Binbin, Chris Brunsdon, Martin Charlton, and Paul Harris. 2017.
{``Geographically Weighted Regression with Parameter-Specific Distance
Metrics.''} \emph{International Journal of Geographical Information
Science} 31 (5): 982--98.

\leavevmode\vadjust pre{\hypertarget{ref-lu2016minkowski}{}}%
Lu, Binbin, Martin Charlton, Chris Brunsdon, and Paul Harris. 2016.
{``The Minkowski Approach for Choosing the Distance Metric in
Geographically Weighted Regression.''} \emph{International Journal of
Geographical Information Science} 30 (2): 351--68.

\leavevmode\vadjust pre{\hypertarget{ref-lu2014gwmodel}{}}%
Lu, Binbin, Paul Harris, Martin Charlton, and Chris Brunsdon. 2014.
{``The GWmodel r Package: Further Topics for Exploring Spatial
Heterogeneity Using Geographically Weighted Models.''} \emph{Geo-Spatial
Information Science} 17 (2): 85--101.

\leavevmode\vadjust pre{\hypertarget{ref-lu2018improvements}{}}%
Lu, Binbin, Wenbai Yang, Yong Ge, and Paul Harris. 2018. {``Improvements
to the Calibration of a Geographically Weighted Regression with
Parameter-Specific Distance Metrics and Bandwidths.''} \emph{Computers,
Environment and Urban Systems} 71: 41--57.

\leavevmode\vadjust pre{\hypertarget{ref-mcmillen2013package}{}}%
McMillen, Daniel, and Maintainer Daniel McMillen. 2013. {``Package
{`McSpatial'}.''} \emph{Nonparametric Spatial Data Analysis. August} 4.

\leavevmode\vadjust pre{\hypertarget{ref-murakami2015area}{}}%
Murakami, Daisuke, and Morito Tsutsumi. 2015. {``Area-to-Point Parameter
Estimation with Geographically Weighted Regression.''} \emph{Journal of
Geographical Systems} 17 (3): 207--25.

\leavevmode\vadjust pre{\hypertarget{ref-murakami2020scalable}{}}%
Murakami, Daisuke, Narumasa Tsutsumida, Takahiro Yoshida, Tomoki Nakaya,
and Binbin Lu. 2020. {``Scalable GWR: A Linear-Time Algorithm for
Large-Scale Geographically Weighted Regression with Polynomial
Kernels.''} \emph{Annals of the American Association of Geographers},
1--22.

\leavevmode\vadjust pre{\hypertarget{ref-nakaya2014gwr4}{}}%
Nakaya, Tomoki, M Charlton, P Lewis, C Brunsdon, J Yao, and S
Fotheringham. 2014. {``Gwr4 User Manual.''} \emph{Windows Application
for Geographically Weighted Regression Modelling}.

\leavevmode\vadjust pre{\hypertarget{ref-openshaw1996developing}{}}%
Openshaw, Stan. 1996. {``Developing GIS-Relevant Zone-Based Spatial
Analysis Methods.''} \emph{Spatial Analysis: Modelling in a GIS
Environment}, 55--73.

\leavevmode\vadjust pre{\hypertarget{ref-oshan2019mgwr}{}}%
Oshan, Taylor M, Ziqi Li, Wei Kang, Levi J Wolf, and A Stewart
Fotheringham. 2019. {``Mgwr: A Python Implementation of Multiscale
Geographically Weighted Regression for Investigating Process Spatial
Heterogeneity and Scale.''} \emph{ISPRS International Journal of
Geo-Information} 8 (6): 269.

\leavevmode\vadjust pre{\hypertarget{ref-paez2002general}{}}%
Páez, Antonio, Takashi Uchida, and Kazuaki Miyamoto. 2002a. {``A General
Framework for Estimation and Inference of Geographically Weighted
Regression Models: 1. Location-Specific Kernel Bandwidths and a Test for
Locational Heterogeneity.''} \emph{Environment and Planning A} 34 (4):
733--54.

\leavevmode\vadjust pre{\hypertarget{ref-paez2002general2}{}}%
---------. 2002b. {``A General Framework for Estimation and Inference of
Geographically Weighted Regression Models: 2. Spatial Association and
Model Specification Tests.''} \emph{Environment and Planning A} 34 (5):
883--904.

\leavevmode\vadjust pre{\hypertarget{ref-pebesma2018simple}{}}%
Pebesma, Edzer J. 2018. {``Simple Features for r: Standardized Support
for Spatial Vector Data.''} \emph{R J.} 10 (1): 439.

\leavevmode\vadjust pre{\hypertarget{ref-pebesma2005s}{}}%
Pebesma, Edzer, and Roger S Bivand. 2005. {``S Classes and Methods for
Spatial Data: The Sp Package.''} \emph{R News} 5 (2): 9--13.

\leavevmode\vadjust pre{\hypertarget{ref-quinones2021geographically}{}}%
Quiñones, Sarah, Aditya Goyal, and Zia U Ahmed. 2021. {``Geographically
Weighted Machine Learning Model for Untangling Spatial Heterogeneity of
Type 2 Diabetes Mellitus (T2d) Prevalence in the USA.''}
\emph{Scientific Reports} 11 (1): 1--13.

\leavevmode\vadjust pre{\hypertarget{ref-rey2010pysal}{}}%
Rey, Sergio J, and Luc Anselin. 2010. {``PySAL: A Python Library of
Spatial Analytical Methods.''} In \emph{Handbook of Applied Spatial
Analysis}, 175--93. Springer.

\leavevmode\vadjust pre{\hypertarget{ref-tobler1970computer}{}}%
Tobler, Waldo R. 1970. {``A Computer Movie Simulating Urban Growth in
the Detroit Region.''} \emph{Economic Geography} 46 (sup1): 234--40.

\leavevmode\vadjust pre{\hypertarget{ref-wheeler2013gwrr}{}}%
Wheeler, David. 2013. {``Gwrr: Fits Geographically Weighted Regression
Models with Diagnostic Tools.''}
\emph{Https://Cran.r-Project.org/Web/Packages/Gwrr/Index.html}.

\leavevmode\vadjust pre{\hypertarget{ref-wheeler2007diagnostic}{}}%
Wheeler, David C. 2007. {``Diagnostic Tools and a Remedial Method for
Collinearity in Geographically Weighted Regression.''} \emph{Environment
and Planning A} 39 (10): 2464--81.

\leavevmode\vadjust pre{\hypertarget{ref-wheeler2009simultaneous}{}}%
---------. 2009. {``Simultaneous Coefficient Penalization and Model
Selection in Geographically Weighted Regression: The Geographically
Weighted Lasso.''} \emph{Environment and Planning A} 41 (3): 722--42.

\leavevmode\vadjust pre{\hypertarget{ref-wickham2014tidy}{}}%
Wickham, Hadley et al. 2014. {``Tidy Data.''} \emph{Journal of
Statistical Software} 59 (10): 1--23.

\leavevmode\vadjust pre{\hypertarget{ref-wickham2019advanced}{}}%
Wickham, Hadley. 2019. \emph{Advanced r}. chapman; hall/CRC.

\leavevmode\vadjust pre{\hypertarget{ref-xu2021spatial}{}}%
Xu, Saiping, Qianjun Zhao, Kai Yin, Guojin He, Zhaoming Zhang, Guizhou
Wang, Meiping Wen, and Ning Zhang. 2021. {``Spatial Downscaling of Land
Surface Temperature Based on a Multi-Factor Geographically Weighted
Machine Learning Model.''} \emph{Remote Sensing} 13 (6): 1186.

\leavevmode\vadjust pre{\hypertarget{ref-yang2014extension}{}}%
Yang, Wenbai. 2014. {``An Extension of Geographically Weighted
Regression with Flexible Bandwidths.''} PhD thesis, University of St
Andrews.

\leavevmode\vadjust pre{\hypertarget{ref-yoneoka2016new}{}}%
Yoneoka, Daisuke, Eiko Saito, and Shinji Nakaoka. 2016. {``New Algorithm
for Constructing Area-Based Index with Geographical Heterogeneities and
Variable Selection: An Application to Gastric Cancer Screening.''}
\emph{Scientific Reports} 6 (1): 1--7.

\end{CSLReferences}

\end{document}